\documentclass[journal=jpclcd,manuscript=article]{achemso}
\setkeys{acs}{keywords = true}
\usepackage[version=3]{mhchem} 
\usepackage[colorlinks=true,allcolors=blue]{hyperref}
\usepackage{amssymb}
\usepackage{amsmath}
\usepackage{graphicx}
\usepackage[caption=false]{subfig}
\usepackage{amsfonts}
\usepackage{xcolor}
\usepackage{epsfig}
\usepackage{color}
\usepackage{bm}
\usepackage{tabularx}
\usepackage{multirow}
\usepackage{mathtools}
\usepackage{xfrac}
\usepackage{blkarray}
\usepackage{bbold}
\usepackage[mathscr]{euscript}
\usepackage{autobreak}
\usepackage{comment}
\usepackage{booktabs}
\usepackage{multirow}
\usepackage{graphicx}
\author{Jijun Huang}
\affiliation
{Beijing Computational Science Research Center, Beijing 100193, China}
\author{Bing Huang}
\email{bing.huang@csrc.ac.cn}
\affiliation
{Beijing Computational Science Research Center, Beijing 100193, China}
\alsoaffiliation
{School of Physics and Astronomy, Beijing Normal University, Beijing 100875, China}
\author{Song Li}
\email{li.song@csrc.ac.cn}
\affiliation
{Beijing Computational Science Research Center, Beijing 100193, China}

\title{Dimensionality-Driven Charge Stabilization of Group-IV Color Centers in Diamond Ultrathin Films}

\begin{document}

\maketitle

\begin{tocentry}

\includegraphics[width=1\columnwidth]{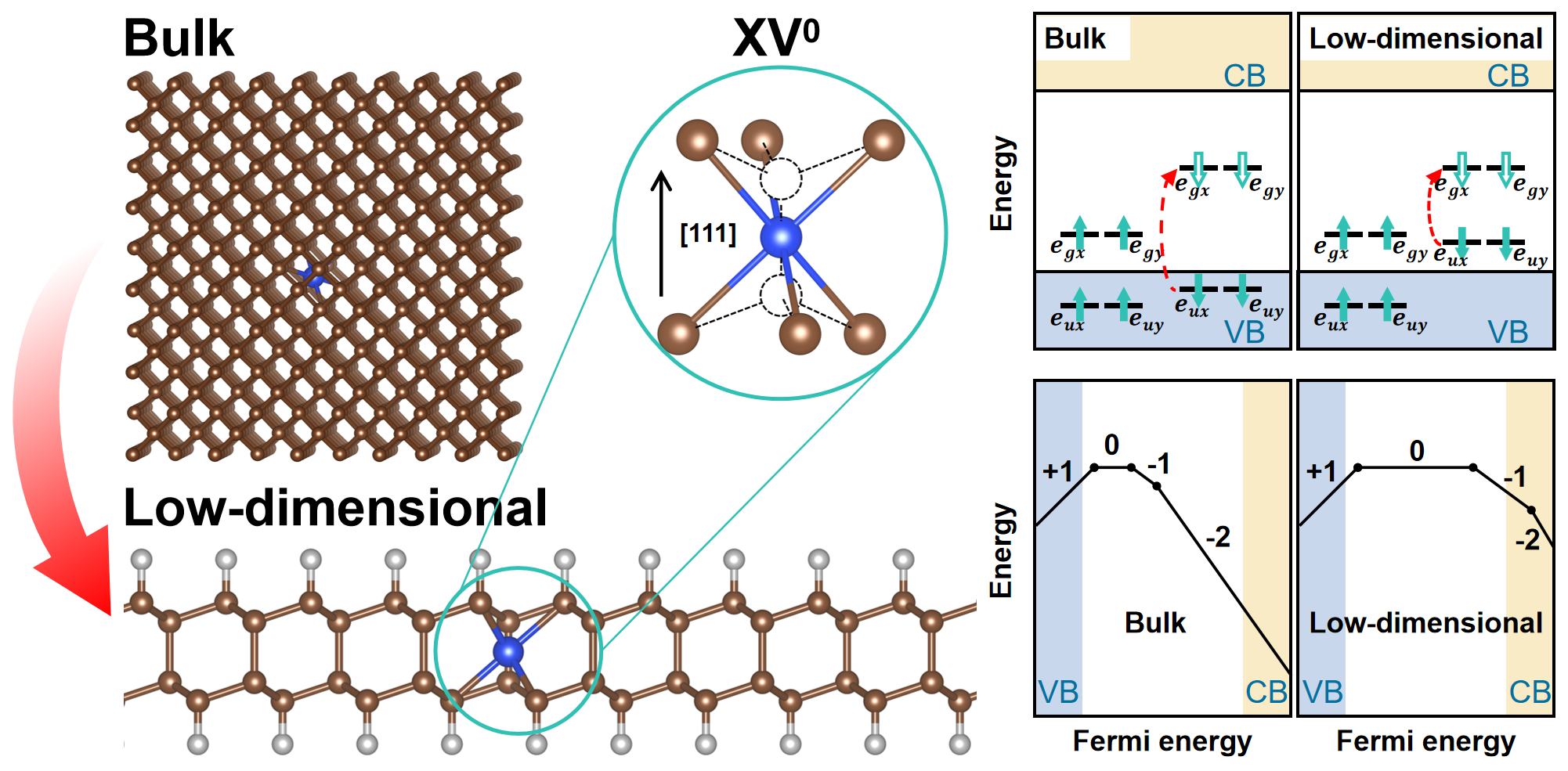}

\end{tocentry}

\begin{abstract}
Neutral group-IV vacancy (XV, X = Si, Ge, Sn, and Pb) centers in diamond are emerging solid-state spin-photon interfaces because of their favorable spin coherence and inversion-symmetry-protected optical transitions. However, stabilizing their neutral charge state typically requires stringent Fermi-level engineering in high-purity boron-doped diamond, which poses significant materials-growth challenges. Here, we demonstrate that dimensional confinement in diamane provides an alternative route to charge-state stabilization without intentional doping. Using first-principles calculations, we show that quantum confinement and surface termination cooperatively tune the host band gap and shift the occupied defect states upward from the valence-band edge, thereby enlarging the thermodynamic stability window of the neutral charge state and suppressing valence-band-assisted excitation pathways. We further reveal that the thickness and surface termination of diamane enable systematic tuning of the electronic structure, zero-field splitting, and spin-orbit coupling of XV centers while largely preserving their optical transition energies. Among the structures considered, hydrogenated diamane offers the most favorable balance between charge-state stability and magneto-optical performance. More broadly, our findings establish dimensional confinement as a general strategy for engineering the charge, optical, and spin properties of solid-state quantum defects.
\end{abstract}

\section*{Keywords}

Diamond; Color centers; Point defects; First-principles calculations; Wide-bandgap semiconductors

\section{Introduction}

The negatively charged nitrogen-vacancy ($\mathrm{NV}^{-}$) center in diamond is one of the most extensively studied solid-state spin defects for quantum sensing~\cite{teissier2014strain,kucsko2013nanometre,maze2008nanoscale,dolde2011electric}and quantum information technologies owing to its optical addressability, long spin coherence time~\cite{balasubramanian2009ultralong}, and operation under ambient conditions~\cite{schirhagl2014nitrogen,childress2013diamond}. Its spin-triplet ground state exhibits a zero-field splitting (ZFS) of 2.87 GHz and can be efficiently initialized, manipulated, and read out through spin-dependent optical transitions. Despite these advantages, the $\mathrm{NV}^{-}$ center suffers from weak zero-phonon-line (ZPL) emission and pronounced spectral diffusion, which limit its performance in photon-mediated quantum technologies.

Group-IV vacancy (XV, X = Si, Ge, Sn, and Pb) centers have emerged as attractive alternatives because of their inversion symmetry, which suppresses coupling to local electric-field fluctuations and enables highly stable optical transitions~\cite{bradac2019quantum,de2021investigation}. In particular, the $\mathrm{SiV}^{-}$ center exhibits a narrow inhomogeneous linewidth and a large Debye–Waller factor, with nearly 70\% of its emission concentrated in the ZPL~\cite{rogers2014multiple}. These properties enable the generation of highly indistinguishable photons and have established XV centers as leading candidates for distributed quantum networks~\cite{stas2022robust,stas2026entanglement}. However, the spin coherence of negatively charged XV centers is often limited by phonon-assisted spin dephasing arising from strong spin–orbit coupling~\cite{sukachev2017silicon}. Recent studies have shown that the neutral charge state can substantially improve spin coherence while preserving the inversion-symmetry-protected optical properties~\cite{green2017neutral}. Similar opportunities are expected for GeV, SnV, and PbV centers, whose larger spin–orbit interactions provide access to quantum operations at elevated temperatures~\cite{cheng2025laser,iwasaki2017tin,guo2023microwave}. Nevertheless, efficient stabilization of the neutral charge state remains a major challenge, and experimental observations of neutral group-IV vacancy centers are still scarce.

Charge-state instability represents a broader obstacle for XV-based quantum technologies. Under resonant optical excitation, even the negatively charged state can undergo photoionization and convert into optically inactive charge states, e.g. $\mathrm{SiV}^{-}$ to $\mathrm{SiV}^{-2}$, resulting in fluorescence loss and degraded device performance~\cite{rieger2025mitigating,rieger2024fast}. Current strategies rely primarily on Fermi-level engineering through boron doping or surface modification to stabilize the desired charge state~\cite{rose2018observation,zhang2023neutral}. While effective in some cases, these approaches introduce additional materials complexity and may alter the optical and spin properties of the defect~\cite{nagl2015improving,rodgers2021materials}. In addition, the ultrahigh hardness of diamond makes micro/nanofabrication and controllable doping difficult. More fundamentally, they do not address whether the intrinsic electronic structure of XV centers can be engineered to favor charge-state stability. Identifying alternative routes to charge-state control that do not rely on heavy doping therefore remains an important challenge.

Low-dimensional diamond structures offer an intriguing opportunity in this context. Recent advances in the synthesis of diamane, an atomically thin diamond film stabilized by surface functionalization, have opened a new platform for defect engineering beyond bulk diamond~\cite{bakharev2020chemically,sorokin2026properties}. Reduced dimensionality can profoundly modify the electronic structure of wide-bandgap semiconductors through quantum confinement and surface electrostatic effects. However, it remains unknown whether such effects can be exploited to systematically engineer the charge-state stability and magneto-optical properties of XV centers.

Here, we demonstrate from first-principles calculations that dimensional confinement in diamane provides an alternative route to charge-state engineering of XV centers without intentional doping. We show that quantum confinement and surface termination cooperatively shift the occupied defect states away from the valence-band edge, thereby enlarging the thermodynamic stability window of the neutral charge state and suppressing valence-band-assisted excitation pathways. In addition, the thickness and surface chemistry of diamane enable systematic tuning of the electron–phonon coupling, zero-field splitting, and spin–orbit interaction while largely preserving the ZPL energies. Our results establish dimensional confinement as a general strategy for controlling the charge, optical, and spin properties of solid-state quantum defects.

\section{Results and discussion}

\subsection{Structural and electronic properties of XV centers}

\begin{figure}
\includegraphics[width=\columnwidth]{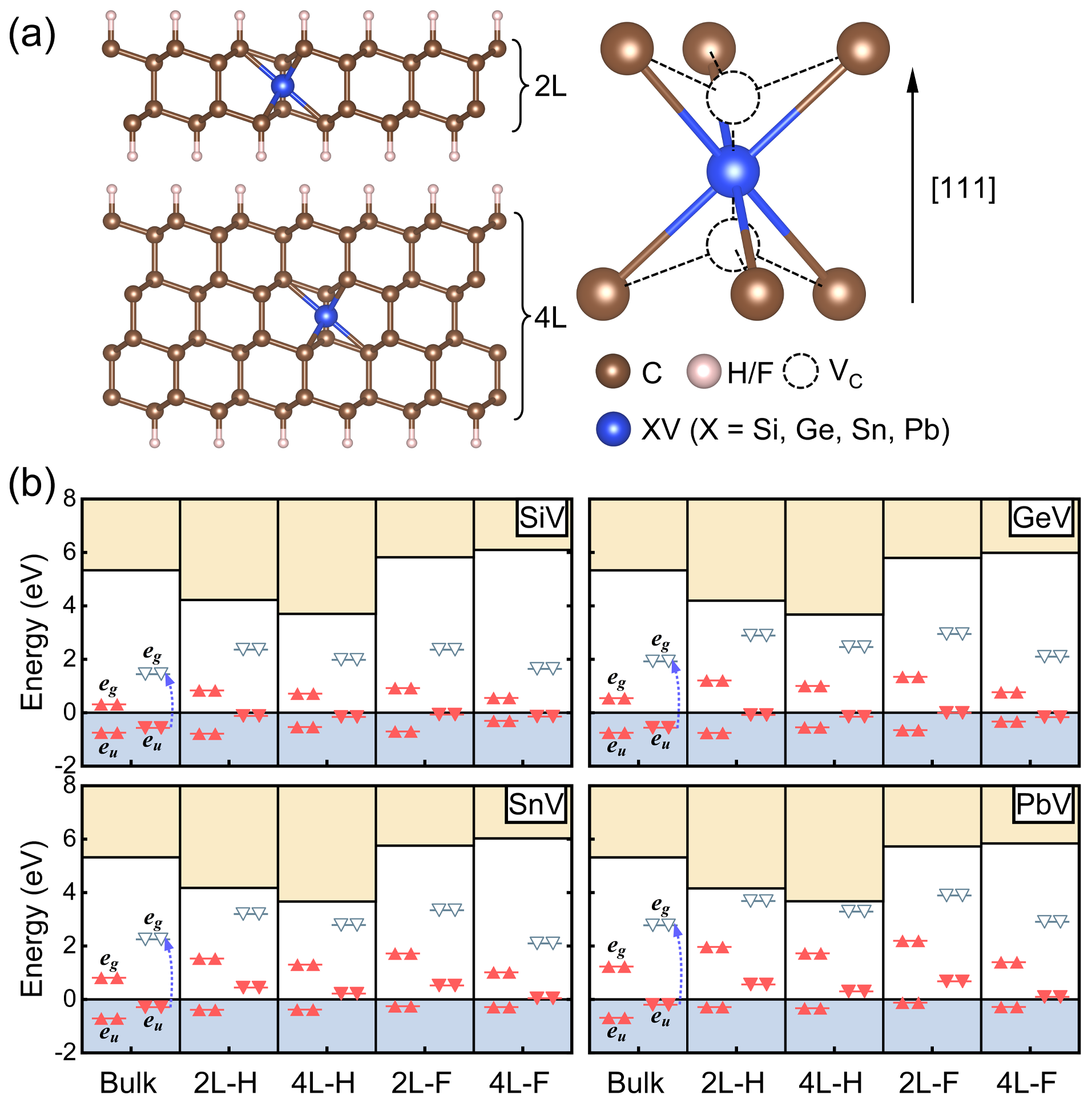}
\caption{\label{fig:1}%
Geometric structure (a) and defect levels (b) of XV centers in H- and F-terminated diamond films along the (111) direction. The defect levels are calculated using the HSE functional. The energy levels are aligned to the VBM of diamane as illustrated by blue. The yellow region is the valence band with the surface states included. The triangle arrows indicate the spin direction and the orange color indicates the occupied states. The empty states are filled with white. The optical excitation is from $e_u$ to $e_g$ orbitals in spin-down channel. Defect levels located inside the band gap are less susceptible to hybridization with the valence-band states during optical excitation, and they are more favorable for realizing localized and stable optical transitions.}
\end{figure}

Few-layer diamond films (diamane), shown in Fig.~\ref{fig:1}(a) are intrinsically unstable in the absence of surface functionalization and tend to reconstruct into sp$^2$-bonded carbon phases~\cite{kvashnin2014phase,piazza2019low}. Surface termination is therefore essential to kinetically stabilize the sp$^3$-bonded diamane structure~\cite{rajasekaran2013interlayer}. Recent experimental advances have demonstrated the successful synthesis of hydrogen- and fluorine-terminated diamane~\cite{bakharev2020chemically,son2020tailoring}, suggesting that incorporation of XV centers during growth is experimentally feasible.

In this work, we construct $n$L-T diamane supercell models, where $n$ denotes the number of diamond layers and T represents the surface termination species. We focus on (111)-oriented diamane, which naturally aligns the symmetry axis of XV centers perpendicular to the film plane (Fig.~\ref{fig:1}). We consider H-, F-, N-, and OH-terminations, which represent experimentally relevant surface chemistries for diamond-based materials~\cite{janitz2022diamond,pakornchote2020roles,kaviani2014proper}. 

The electronic structure of XV centers is found to be governed by a cooperative interplay between quantum confinement and surface electrostatics. Quantum confinement primarily determines the band-gap and defect-state reorganization, whereas surface termination controls the electrostatic boundary condition and electron affinity, thereby tuning the relative alignment between defect levels and host band edges. As a consequence, the energetic position of the occupied spin-down $e_u$ states is highly sensitive to both thickness and termination.

We find that H- and F-terminated diamane systematically shift the $e_u$ levels of SiV centers upward relative to bulk diamond by up to 0.425 eV (4L-F), effectively decreasing their energetic separation from the valence-band maximum (VBM). In contrast, N- and OH-terminations induce a downward shift of up to 0.999 eV due to surface-state-induced band-edge pinning, which enhances coupling between the $e_u$ levels and the valence band and thus promotes undesired valence-band-assisted excitation pathways.

Importantly, this behavior directly correlates with charge-state stability: decoupling of the $e_u$ levels from the VBM suppresses resonant excitation processes and favors stabilization of the neutral charge state. Based on this mechanism, H- and F-terminated diamane emerge as the most suitable platforms for increasing the emission purity.

Structural relaxations indicate that the local geometry of XV centers is largely preserved upon dimensional confinement (Fig. S3). The C-X bond lengths exhibit only minor variations with thickness, and the characteristic $D_{3d}$ symmetry of the defect is maintained. In the ultrathin limit (2L), surface hydrogen atoms exhibit slight inward relaxation toward the defect site, while this distortion is progressively suppressed with increasing thickness. Despite these surface relaxations, the XV centers retained the characteristic D$_3d$ symmetry.

Fig.~\ref{fig:1}(b) shows the HSE levels of XV centers in H- and F-terminated diamane. For clarity, only the doubly degenerate e$_u$ and e$_g$ levels located inside or near the band gap are shown. In the neutral charge state, the e$_u$ levels are fully occupied, whereas the e$_g$ levels are occupied only in the spin-up channel and remain empty in the spin-down channel. The two unpaired electrons thus give rise to a spin-triplet ground state ($S = 1$). Both H- and F-terminated diamane induce a general upward shift of the spin-down $e_u$ levels. For SiV and GeV centers, these states remain close to the VBM, whereas for SnV and PbV centers they enter the band gap and become effectively decoupled from the valence band. The degree of decoupling increases with thickness, as illustrated by the reduction of the $e_u$-VBM separation from 0.444 eV (2L-H) to 0.216 eV (4L-H) and from 0.523 eV (2L-F) to 0.045 eV (4L-F) in SnV centers. Similar trends are observed across all XV species.

Overall, these results establish that dimensional confinement and surface termination provide two coupled but distinct tuning parameters for engineering the electronic structure of XV centers in diamane, enabling systematic control over defect-level and host-band interactions for elevated temperature operation.


\subsection{Defect formation energies of XV centers}

\begin{figure}
\includegraphics[width=\columnwidth]{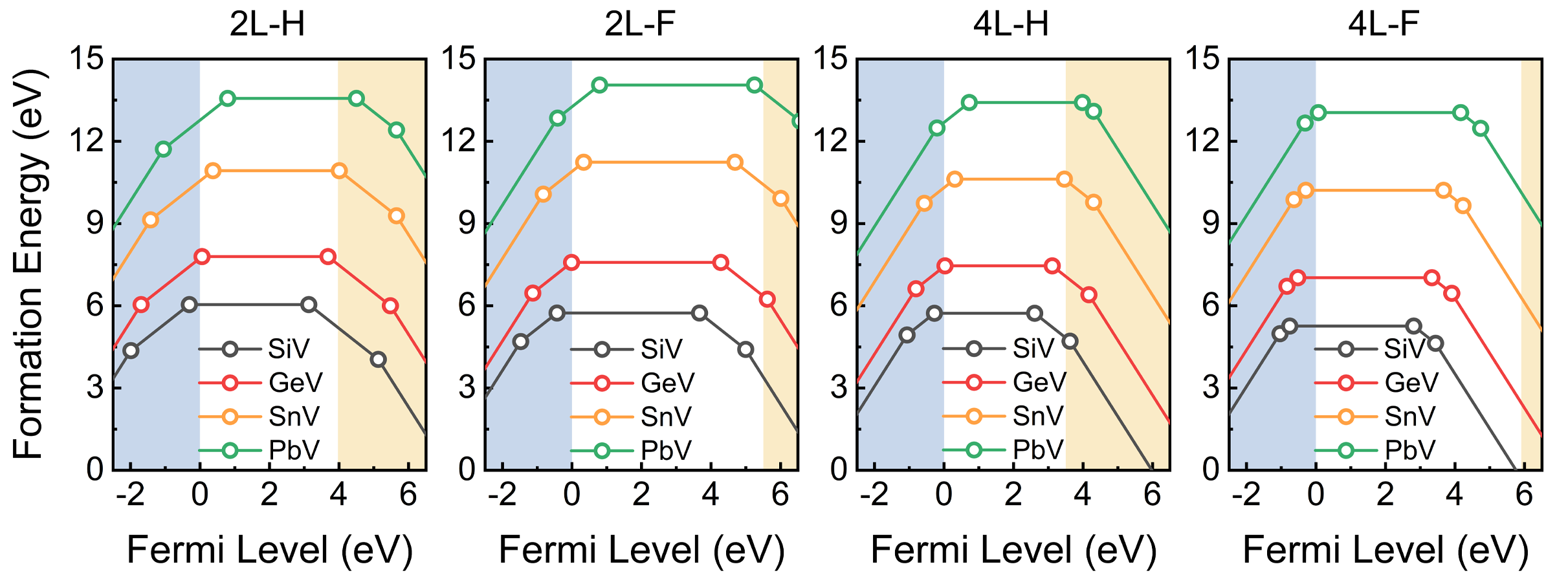}
\caption{\label{fig:2}%
Defect formation energies of XV centers in H- and F-terminated diamond films. The formation energies of XV centers follow the trend E$_f$ (PbV) $>$ E$_f$(SnV) $>$ E$_f$(GeV) $>$ E$_f$(SiV).}
\end{figure}

To determine whether dimensional confinement can stabilize neutral group-IV vacancy centers, we calculated the formation energies and charge-transition levels (CTLs) of XV defects in diamane, as summarized in Fig.~\ref{fig:2}. A striking feature is the substantial expansion of the thermodynamic stability window of the neutral charge state compared with bulk diamond. In contrast to bulk systems, where stabilization of neutral XV centers generally requires Fermi-level engineering through heavy boron doping, diamane intrinsically favors the neutral charge state over a broad Fermi-level range.

The enhanced stability originates from the confinement-induced shift of the defect electronic structure discussed above. The upward shift of the occupied $e_u$ levels weakens their interaction with the valence-band states and drives the relevant CTLs toward the conduction-band minimum (CBM), thereby enlarging the energetic window in which the neutral charge state is thermodynamically favored. This effect becomes progressively stronger with increasing atomic number of the group-IV impurity, leading to systematically improved stabilization of neutral GeV, SnV, and PbV centers.

Surface termination further modulates the charge-state landscape. In H-terminated diamane, the neutral charge state already becomes significantly more accessible than in bulk diamond, although the singly negative SiV state remains stable within the band gap. In F-terminated diamane, the larger host band gap further expands the available charge-state-engineering window and enhances the stability of the negative charge states. As a result, both neutral and negatively charged configurations can be selectively stabilized through Fermi-level tuning or optical excitation.

Film thickness provides an additional degree of control. As the thickness increases from 2L to 4L, the CTLs gradually shift toward the valence-band edge and recover the behavior characteristic of bulk diamond~\cite{thiering2018ab}. Consequently, the confinement effect becomes weaker in thicker films. In 4L-F diamane, the doubly negative charge state enters the band gap and reduces the stability range of the singly negative charge state. Nevertheless, these charge states remain optically addressable because photoionization into the phonon sideband can efficiently convert them back to the neutral configuration under visible or blue-light excitation. The detailed CTLs are listed in Table S1.

\subsection{Magneto-optical properties of XV centers}

\begin{figure}
\includegraphics[width=\columnwidth]{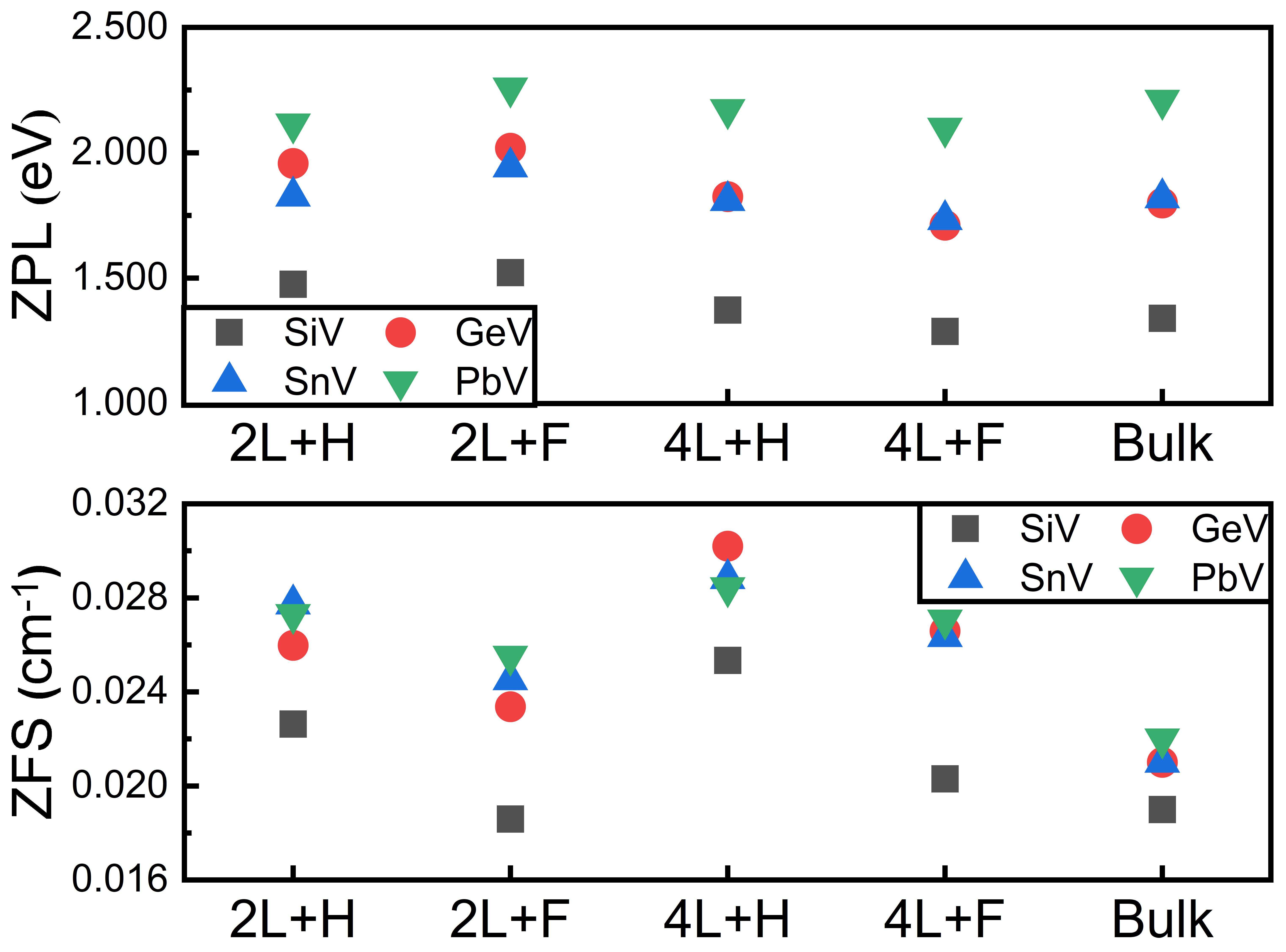}
\caption{\label{fig:3}%
ZPL energies and ZFS parameters of XV centers in H-terminated, F-terminated, and bulk diamond. Only the D$^{SS}$ part is added.}
\end{figure}

The optical excitation considered here corresponds to the promotion of an electron from the occupied spin-down e$_u$ state to the empty e$_g$ state shown in Fig.~\ref{fig:1}. Upon excitation, the partially occupied degenerate orbitals give rise to a product Jahn–Teller (JT) effect, resulting in strong coupling between the electronic states and E$_g$ vibrational modes, described by the $(e_u\otimes e_g)\otimes E_g$ interaction~\cite{thiering2019eg}. Consequently, the excited-state geometry relaxes from the high-symmetry D${3d}$ configuration to a lower-symmetry C${2h}$ structure. Here we focus on the lowest optically active transition and its associated ZPL, summarized in Fig.~\ref{fig:3} and Table~\ref{tab:zpl_hr_dw_lifetime_zfs}. Despite substantial changes in charge-state stability and defect-level alignment, the calculated ZPL energies remain remarkably close to their bulk counterparts. So the conventional 532 nm continuous-wave laser can still be used to measure the photoluminescence spectra. Among the XV centers, the ZPL energy increases from SiV to PbV. A slight redshift of the ZPL is observed with increasing film thickness. 

\begin{figure}
\includegraphics[width=\columnwidth]{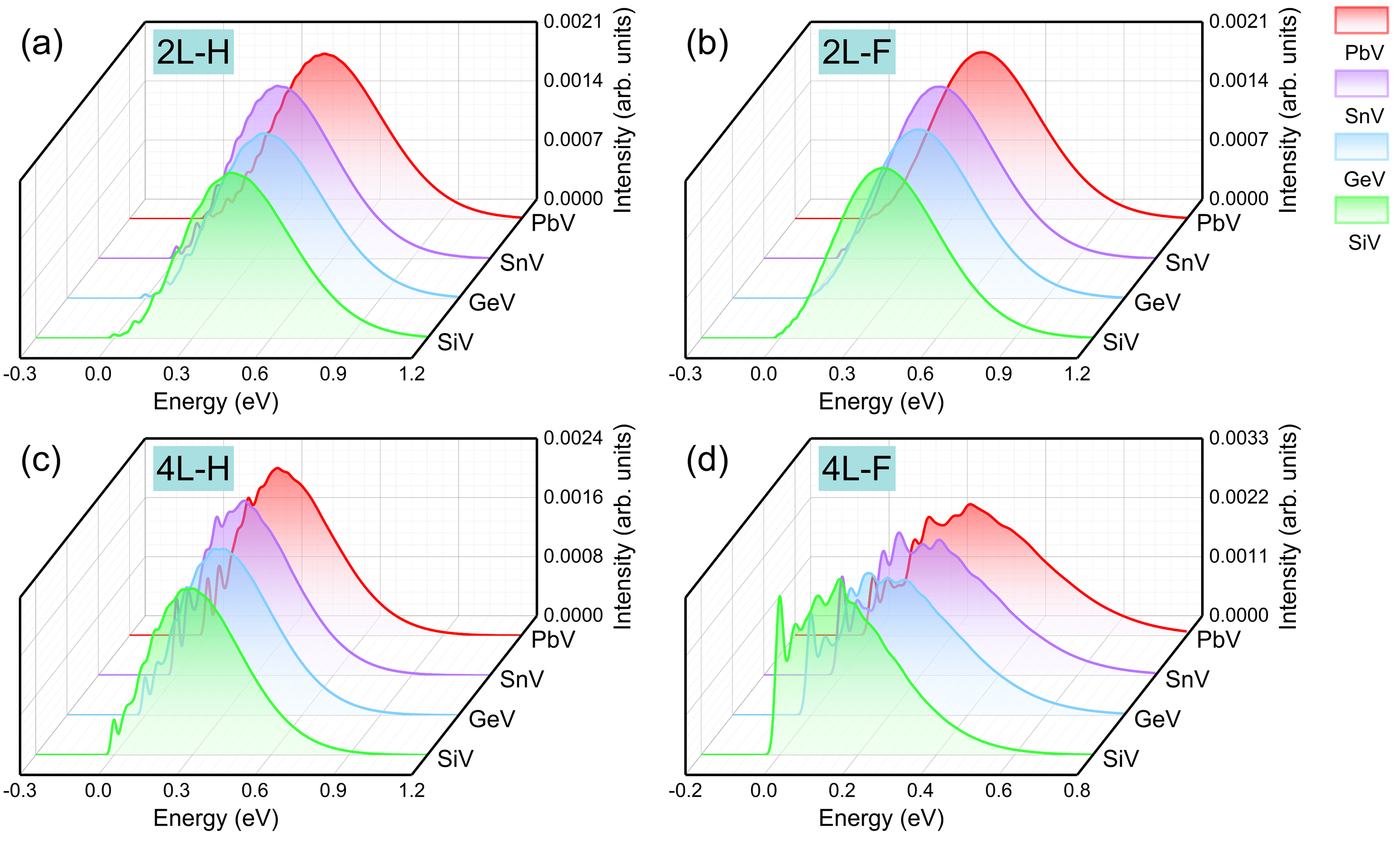}
\caption{\label{fig:4}%
Simulated PL spectra of XV centers in H- and F-terminated diamond films. The energies are aligned to ZPL.}
\end{figure}

In contrast to the weak dependence of the ZPL energies, the electron–phonon coupling exhibits a pronounced sensitivity to dimensional confinement and surface termination, as listed in Tab.~\ref{tab:zpl_hr_dw_lifetime_zfs}. In the ultrathin limit, the Huang–Rhys (HR) factor reaches 5.498, resulting in a small Debye–Waller (DW) factor of only 0.4\%. The optical emission is therefore dominated by phonon sidebands rather than direct ZPL emission. The enhanced electron–phonon coupling originates from the pronounced excited-state lattice relaxation induced by surface effects in ultrathin diamane. The calculated Jahn-Teller energies (see Tab. S2) further indicate that the excited states reside in the static Jahn–Teller regime. Increasing the film thickness progressively suppresses surface-induced distortions and weakens the electron–phonon interaction. Consequently, the DW factor increases to as high as 7.3\% in 4L-F SiV, representing the most favorable optical performance among the systems considered. The calculated photoluminescence spectra are fully consistent with the HR and DW factors discussed above (Fig.~\ref{fig:4}a,b). Strong electron–phonon coupling in ultrathin diamane leads to phonon-sideband-dominated emission, whereas increasing the film thickness progressively restores ZPL emission. 

To further isolate the role of the product Jahn–Teller effect, we calculated the optical properties of the SiV center in 2L-H while constraining the excited-state geometry to the high-symmetry configuration. As shown in Fig. S4, the HR factor decreases dramatically to 0.380, accompanied by a substantial suppression of the phonon sideband. Similar trends are observed for the other XV centers. These results confirm that the strong electron–phonon coupling in ultrathin diamane originates primarily from the product Jahn–Teller distortion and highlight the importance of explicitly accounting for this effect when modeling excited states with partially occupied degenerate orbitals.

The calculated radiative lifetimes decrease systematically with increasing atomic number and film thickness (Table~\ref{tab:zpl_hr_dw_lifetime_zfs}). In thicker diamane structures, the lifetimes approach those reported for bulk XV centers~\cite{thiering2018ab}, indicating enhanced radiative transition rates and improved optical performance. Together, these results reveal a clear trade-off between charge-state stabilization and optical coherence in ultrathin diamane. While strong confinement substantially enhances the stability of the neutral charge state, it simultaneously amplifies product Jahn–Teller distortions and electron–phonon coupling. Increasing the film thickness partially restores bulk-like optical behavior while retaining the benefits of confinement-induced charge-state engineering.

\begin{table}[htbp]
    \centering
    \caption{ZPL energy (eV), HR factor, DW factor, radiative lifetime (ns), and ZFS parameter \(D^{\mathrm{SS}}\) (cm\(^{-1}\)) of XV centers in H- and F-terminated diamond films.}
    \label{tab:zpl_hr_dw_lifetime_zfs}
    \resizebox{\textwidth}{!}{
    \begin{tabular}{llccccc}
        \toprule
        System & Defect & ZPL & HR & DW & Radiative lifetime & \(D^{\mathrm{SS}}\) \\
        \midrule

        \multirow{4}{*}{2L-H} 
        & SiV & 1.475 & 6.630 & 0.001 & 44.163 & 0.023 \\
        & GeV & 1.957 & 6.530 & 0.002 & 23.664 & 0.026 \\
        & SnV & 1.826 & 5.498 & 0.004 & 31.552 & 0.028 \\
        & PbV & 2.115 & 6.121 & 0.002 & 24.667 & 0.027 \\
        \midrule

        \multirow{4}{*}{2L-F} 
        & SiV & 1.521 & 6.585 & 0.001 & 36.917 & 0.019 \\
        & GeV & 2.108 & 7.033 & 0.001 & 20.359 & 0.023 \\
        & SnV & 1.943 & 5.883 & 0.003 & 25.080 & 0.025 \\
        & PbV & 2.259 & 6.711 & 0.001 & 19.866 & 0.026 \\
        \midrule

        \multirow{4}{*}{4L-H} 
        & SiV & 1.373 & 4.367 & 0.013 & 3.828 & 0.025 \\
        & GeV & 1.825 & 4.290 & 0.014 & 2.078 & 0.030 \\
        & SnV & 1.809 & 3.599 & 0.027 & 2.360 & 0.029 \\
        & PbV & 2.173 & 3.880 & 0.021 & 1.669 & 0.028 \\
        \midrule

        \multirow{4}{*}{4L-F} 
        & SiV & 1.288 & 2.620 & 0.073 & 4.698 & 0.020 \\
        & GeV & 1.711 & 3.047 & 0.048 & 2.821 & 0.027 \\
        & SnV & 1.733 & 3.053 & 0.047 & 2.796 & 0.026 \\
        & PbV & 2.098 & 3.549 & 0.029 & 1.994 & 0.027 \\
        \midrule

        \multirow{4}{*}{Bulk} 
        & SiV & 1.34~\cite{thiering2019eg} & 1.5~\cite{PhysRevB.84.245208} & -- & -- & 0.019~\cite{PhysRevResearch.2.023071} \\
        & GeV & 1.80~\cite{thiering2019eg} & -- & -- & -- & 0.021~\cite{PhysRevResearch.2.023071} \\
        & SnV & 1.82~\cite{thiering2019eg} & -- & -- & -- & 0.021~\cite{PhysRevResearch.2.023071} \\
        & PbV & 2.21~\cite{thiering2019eg} & -- & -- & -- & 0.022~\cite{PhysRevResearch.2.023071} \\
        \bottomrule
    \end{tabular}
    }
\end{table}

The calculated ZFS parameters of neutral XV centers are summarized in Fig.~\ref{fig:3} and Table~\ref{tab:zpl_hr_dw_lifetime_zfs}. In contrast to the $\mathrm{NV}^{-}$ center, whose ZFS is dominated by the spin–spin dipolar interaction, the ZFS of XV centers is governed primarily by the second-order spin–orbit contribution ($D^{SOC}$). As a result, the overall ZFS increases substantially with the atomic number of the group-IV centers. The spin–spin component ($D^{SS}$) remains comparatively insensitive to both thickness and termination because the spin-density distributions are similar across the XV family. The calculated $D^{SS}$ values are only slightly larger than those reported for bulk diamond. The large SOC-induced ZFS is advantageous for suppressing phonon-mediated spin decoherence and enables spin manipulation at elevated temperatures.

\begin{figure}
\includegraphics[width=\columnwidth]{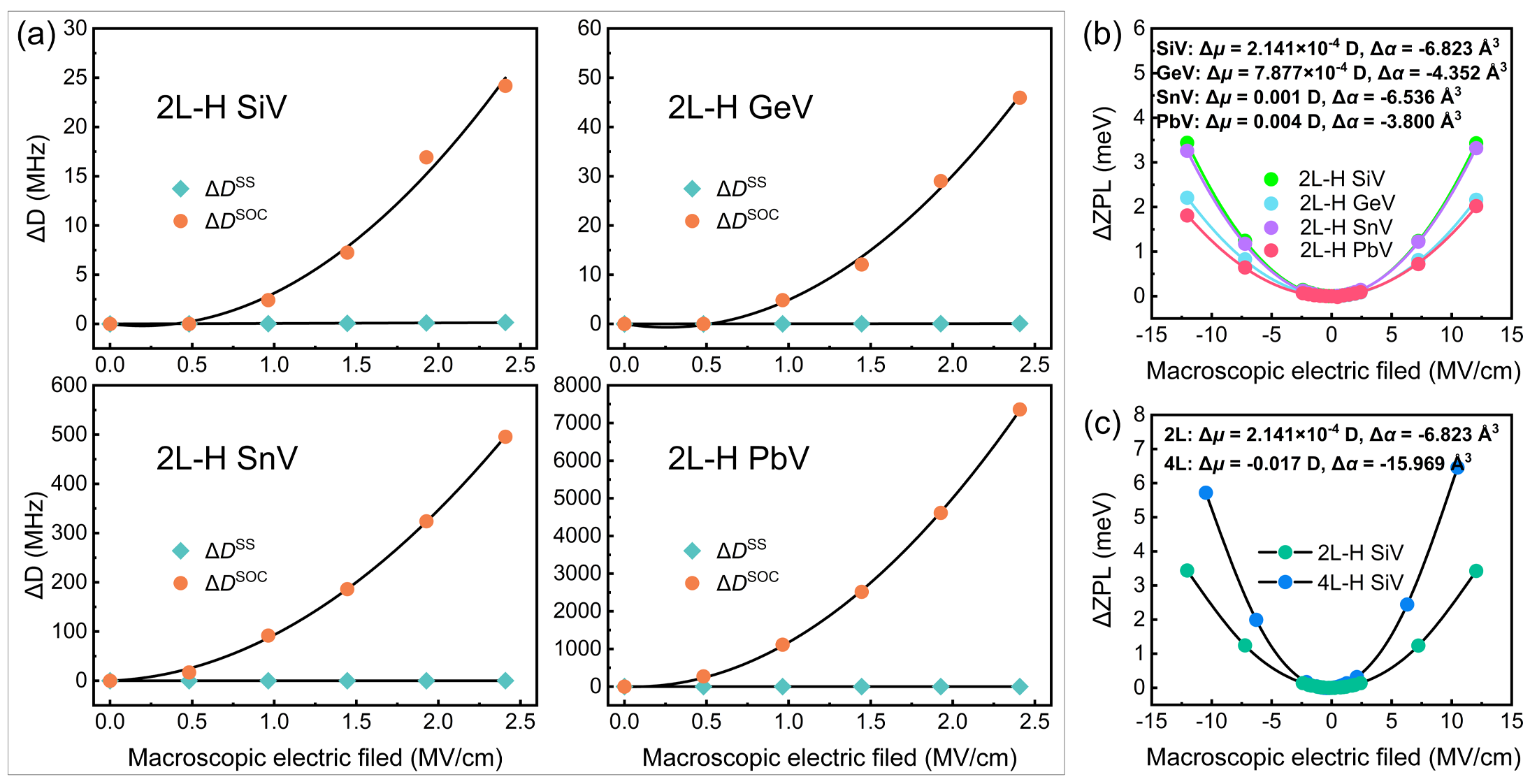}
\caption{\label{fig:5}%
(a) Stark shifts of the ZFS for XV centers in 2L-H relative to the case without an electric field. In 2L-H, the Stark coefficients are: 0.051 GHz/(V/\AA)$^{-2}$ in SiV, 0.102 GHz/(V/\AA)$^{-2}$ in GeV, 8.020 GHz/(V/\AA)$^{-2}$ in SnV, and 13.289 GHz/(V/\AA)$^{-2}$ in PbV. (b) Stark shifts of the ZPL of XV centers in 2L-H; (c) Stark shifts of ZPL of SiV centers in 2L-H and 4L-H, with energy offsets referenced to the ZPL in the absence of an electric field.}
\end{figure}

The ZFS exhibits a pronounced electric-field dependence that is almost entirely mediated by the spin–orbit interaction. As shown in Fig.~\ref{fig:5}a, while the spin–spin contribution remains essentially unchanged, the SOC component increases strongly under an applied field. The field dependence follows a quadratic behavior and the extracted Stark coefficients increase dramatically with atomic number, ranging from 0.051 GHz/(V/\AA)$^{-2}$ for SiV to 13.289 GHz/(V/\AA)$^{-2}$ for PbV. This trend reveals a substantial enhancement of electric-field tunability in heavier group-IV vacancy centers.

The Stark response of the optical transition is also evaluated in Fig.~\ref{fig:5}. The calculated change in dipole moment ($\Delta\mu$) remains extremely small, rendering the polarizability difference ($\Delta\alpha$) the dominant contribution to the ZPL shift. Accordingly, the field dependence is well described by a quadratic Stark model,

\[
\Delta E_{ZPL}
=
-\Delta\mu\mathbf{E}
-\frac{1}{2}\mathbf{E}\Delta\alpha\mathbf{E}.
\]

For all XV centers, the extracted $\Delta\mu$ values remain several orders of magnitude smaller than those of the $\mathrm{NV}^{-}$ center. For example, $\Delta\mu$ is only 0.001 D for SnV in 2L-H diamane, despite being approximately five times larger than the corresponding bulk value~\cite{aghaeimeibodi2021electrical}. The calculated $\Delta\alpha$ values are negative, indicating that the ground state is more polarizable than the excited state, consistent with previous observations for $\mathrm{NV}^{-}$ centers~\cite{tamarat2006stark}. Although $\Delta\mu$ increases monotonically with atomic number, the variation of $\Delta\alpha$ is non-monotonic. In particular, SnV exhibits a larger polarizability change than GeV, a trend that parallels the evolution of the ZPL energies and may originate from enhanced $p–d$ orbital hybridization~\cite{qiu2023origin}.

\section{Discussion}

The experimental feasibility of this strategy is supported by a recent report demonstrating the incorporation of SiV and GeV centers into hydrogen-terminated nanodiamonds with diameters of only 3–4 nm, without ion implantation, irradiation, or post-growth processing~\cite{liang2026bottom}. Interestingly, singly negative rather than doubly negative charge states were observed, consistent with the confinement-induced charge-state reorganization predicted here. These observations suggest that nanoscale dimensional confinement can substantially modify the charge-state landscape of group-IV vacancy centers. Further reduction of the crystal size may therefore provide a route toward stabilizing the neutral charge state.

Beyond charge-state stabilization, diamane offers several attractive features for quantum-device integration. Its atomically thin geometry naturally places defect centers in close proximity to external targets, making it particularly appealing for nanoscale sensing and hybrid quantum architectures. At the same time, our calculations reveal an intrinsic trade-off associated with strong dimensional confinement. While ultrathin diamane substantially improves the stability of neutral XV centers, it also enhances electron–phonon coupling through surface-induced Jahn–Teller distortions, leading to reduced Debye–Waller factors and less favorable photon indistinguishability. Importantly, this trade-off can be mitigated through thickness engineering. Increasing the film thickness progressively suppresses surface distortions while retaining the confinement-induced enhancement of charge-state stability. Indeed, the enlarged neutral-state stability window persists even in the 8L structures terminated by hydrogen or oxygen (Fig. S5), despite oxygen termination generally being unfavorable for neutral-state stabilization in bulk diamond. A remaining challenge is the potential reduction of spin coherence arising from nuclear-spin-rich surface terminations such as hydrogen~\cite{ryan2018impact}. One possible solution is isotopic surface engineering using deuterium, which eliminates the nuclear-spin bath associated with surface hydrogen~\cite{li2025non}.

\section{Conclusion}

In summary, we investigated neutral group-IV vacancy centers in H- and F-terminated diamane and revealed how dimensional confinement and surface chemistry jointly govern their charge, optical, and spin properties. Quantum confinement reorganizes the defect-level position, substantially enlarging the thermodynamic stability window of the neutral charge state while suppressing its interaction with valence-band states. At the same time, the optical properties can be systematically tuned through film thickness, which weakens electron–phonon coupling and progressively restores bulk-like emission characteristics. These results highlight a controllable trade-off between charge-state stability and optical performance and provide practical design guidelines for defect engineering in low-dimensional diamond materials. More broadly, our work identifies dimensional confinement as an alternative strategy for charge-state engineering of solid-state quantum defects beyond conventional approaches based on Fermi-level control.

\section{Materials and methods}
\section{First-principles calculations}
All first-principles calculations in this work were performed using the Vienna Ab initio Simulation Package (VASP)~\cite{kresse1996efficiency,kresse1996efficient}. The interactions between valence electrons and ionic cores were described using the projector augmented-wave (PAW) method~\cite{blochl1994projector,kresse1999ultrasoft}, and we set a plane-wave cutoff energy of 400 eV for all calculations. Both structural relaxations and electronic-structure calculations were carried out using the Heyd-Scuseria-Ernzerhof (HSE) hybrid functional, which provides a more reliable description of the band gap and defect levels ~\cite{heyd2003hybrid}. The Brillouin zone was sampled at the single Gamma point. The convergence criteria for the total energy and atomic forces were set to 1 $\times$ 10$^{-4}$ eV and 0.01 eV/\AA, respectively. Excited state calculations were performed using the $\Delta$SCF method, which has been widely used to obtain reliable ZPL energies~\cite{gali2023recent,li2026computation}. A 6 $\times$ 6 diamond thin-film supercell was constructed and a vacuum region of 10 \AA\ was introduced along the surface-normal direction to suppress interactions between periodic images. To saturate the surface dangling bonds and avoid artificial dipole interaction, the upper and lower surfaces of each slab were passivated with the same termination species. 

The formation energy of an XV center in charge state $q$ was given by~\cite{sfz1-zvdb}

\begin{equation}
\begin{split}
E^q_\text{f} = &E^q_\text{XV} - E_\text{pri} -\sum_i n_i\mu_i + q\left(E^\text{pri}_\text{VBM} + E_\text{Fermi}\right) + E_\text{corr}\left(q\right)\text{,}
\end{split}
\end{equation}

here, $E_{\mathrm{XV}}^q$ and $E_{\mathrm{pri}}$ are the total energies of the defective and pristine supercells, respectively. $\mu_i$ is the chemical potential of element $i$, which was taken from the corresponding elemental reference phase, and $n_i$ is the number of atoms added ($n_i>$0) to or removed ($n_i<$0) from the pristine supercell. $E_{\text{Fermi}}$ is the Fermi level referenced to the valence-band maximum ($E^{\text{pri}}_{\text{VBM}}$) of the pristine supercell, and $E_{\text{corr}}(q)$ is the finite-size charge correction term. The electrostatic correction for charged defects was treated using the Freysoldt-Neugebauer-Van de Walle (FNV) scheme implemented in the SXDEFECTALIGN2D code~\cite{freysoldt2020generalized}.

For spin systems with $S\geq1$, the spin sublevels can split even in the absence of an external magnetic field due to the electron spin-spin dipolar interaction. For the neutral XV centers, the ground state is a spin-triplet state with $S=1$, and the spin-spin Hamiltonian is given by~\cite{PhysRevB.90.235205,PhysRevMaterials.8.056201}
\begin{equation}
\hat{H}_{\rm ss}=D_{xx}^{\rm SS}\hat{S}_x^2+D_{yy}^{\rm SS}\hat{S}_y^2+D_{zz}^{\rm SS}\hat{S}_z^2
=D^{\rm SS}\left(\hat{S}_z^2-\frac{S(S+1)}{3}\right)+\frac{E^{\rm SS}(\hat{S}_+^2+\hat{S}_-^2)}{2}\text{,}
\label{eq:ZFS_ham}
\end{equation}
here, $D_{xx}^{\rm SS},D_{yy}^{\rm SS},$ and $D_{zz}^{\rm SS}$ are the diagonal components of the ZFS tensor. $\hat{S}_x,\hat{S}_y,$ and $\hat{S}_z$ are the spin matrices, with $\hat{S}^2=\hat{S}_x^2+\hat{S}_y^2+\hat{S}_z^2$. $\hat{S}_\pm=\hat{S}_x\pm i\hat{S}_y$ are the spin raising and lowering operator. $D^{\rm SS}$ and $E^{\rm SS}$ are the axial and transverse ZFS parameters, respectively, and are given by\,
\begin{equation}
D^{\rm SS}=\frac{3}{2}D_{zz}^{\rm SS}\quad\text{and}\quad E^{\rm SS}=\frac{D_{yy}^{\rm SS}-D_{xx}^{\rm SS}}{2}
\label{eq:DE_def}
\end{equation}

In addition to the spin-spin dipolar interaction, the SOC contribution to the ZFS was further considered. When SOC is included, the total energy depends on the spin quantization direction because of the SOC-induced magnetic anisotropy. For an axially symmetric system, the SOC-induced magnetic anisotropy energy can be defined as~\cite{PhysRevResearch.2.023071} \,
\begin{equation}
\Delta E^{\rm SOC}
=
E_{\rm tot}^{\rm SOC}(\parallel)
-
E_{\rm tot}^{\rm SOC}(\perp),
\end{equation}
where \(E_{\mathrm{tot}}^{\mathrm{SOC}}(\parallel)\) and 
\(E_{\mathrm{tot}}^{\mathrm{SOC}}(\perp)\) are the total energies obtained from SOC calculations with the spin quantization axis parallel and perpendicular to 
the principal symmetry axis of the defect, respectively. The corresponding SOC contribution to the axial ZFS parameter is related to 
\(\Delta E^{\mathrm{SOC}}\) by~\cite{PhysRevResearch.2.023071,https://doi.org/10.1002/anie.200390099}
\begin{equation}
D^{\mathrm{SOC}}
=
\begin{cases}
\dfrac{\Delta E^{\mathrm{SOC}}}{S^2}, & \text{for integer } S, \\[6pt]
\dfrac{\Delta E^{\mathrm{SOC}}}{S^2-1/4}, & \text{for half-integer } S.
\end{cases}
\label{eq:D_SOC}
\end{equation}
Since the neutral XV centers considered here have a spin-triplet ground state with \(S=1\), the integer-spin expression in Eq.~\eqref{eq:D_SOC} was used. The total axial ZFS parameter can then be written as
\begin{equation}
D^{\mathrm{tot}}
=
D^{\mathrm{SS}}
+
D^{\mathrm{SOC}}.
\label{eq:D_tot}
\end{equation}
For ideal 
\(D_{3d}\)-symmetric XV centers, the total transverse ZFS parameter \(E\) is expected to be zero or negligibly small because of the equivalence of the two directions perpendicular to the defect symmetry axis. Therefore, the ZFS is mainly characterized by \(D^{\mathrm{tot}}\), which serves as an important spectroscopic fingerprint for 
identifying the defect center and estimating the microwave frequency required for spin manipulation.

The radiative lifetime can be evaluated as~\cite{doi:10.1021/acs.jpclett.2c02687} 
\begin{equation}
\frac{1}{\tau_{\rm rad}}=\frac{n_D E_{\rm ZPL}^3 \mu^2}{3\pi \varepsilon_0 c^3 \hbar^4}
\label{eq:rad_life}
\end{equation}

Here, n$_D$ is the refractive index of H- and F-terminated diamond~\cite{bouzidi2021optical}, E$_{ZPL}$ and $\mu_D$ are the ZPL energy and corresponding transition dipole moment, respectively, and $\varepsilon_0$, $c$, and $\hbar$ denote the vacuum dielectric constant, the speed of light, and the reduced Planck constant, respectively. 

\section*{Acknowledgements}

This work is supported by Science
Challenge Project (Grant No. TZ2025013). B.H. acknowledges the NSFC (Grant No. W2511008, 12088101),
the National Key Research and Development of China
(Grant No. 2022YFA1402400), and NSAF (Grant No.
U2230402).

\section*{Supporting information}

The following files are available free of charge.
\begin{itemize}
  \item SI contains the additional theoretical calculation results, including the formation energy and transition level diagram, the phonon side band without Jahn-Teller distortion.

\end{itemize}

\bibliography{name.bib}

\end{document}